\documentclass[10pt]{article}
\usepackage{graphicx}

\begin{document}

\title{Computational time-reversal imaging with a small number of random and noisy measurements}

\author{M. Andrecut}

\date{ }

\maketitle

{\par\centering IBI, University of Calgary \par}
{\par\centering 2500 University Drive NW, Calgary \par}
{\par\centering Alberta, T2N 1N4, Canada \par}

\noindent

\begin{abstract}
Computational time reversal imaging can be used to locate the position of
multiple scatterers in a known background medium. The current methods for
computational time reversal imaging are based on the null subspace
projection operator, obtained through the singular value decomposition of
the frequency response matrix. Here, we discuss the image recovery problem
from a small number of random and noisy measurements, and we show that this
problem is equivalent to a randomized approximation of the null subspace of
the frequency response matrix. 
\end{abstract}

\bigskip 

PACS: 

02.30.Zz Inverse problems  

43.60.Pt Signal processing techniques for acoustic inverse problems  

43.60.Tj Wave front reconstruction, acoustic time-reversal 

\pagebreak

\section{Introduction}

Computational time-reversal imaging (CTRI) has become an important research
area in recent years, with relevant applications in radar imaging,
exploration seismics, nondestructive material testing, medical imaging [1-9]
etc. CTRI uses the information carried by scattered acoustic, elastic or
electro-magnetic waves to obtain images of the investigated domain [1]. It
was shown that scattered acoustic waves can be time-reversed and focused
onto their original source location through arbitrary media, using a
so-called time-reversal mirror [2]. This important result shows how one can
use CTRI to identify the location of multiple point scatterers (targets) in
a known background medium [3]. In this case, a back-propagated signal is
computed, rather than implemented in the real medium, and its peaks indicate
the existence of possible scattering targets. The current methods for CTRI
are based on the null subspace projection operator, obtained through the
singular value decomposition (SVD) of the frequency response matrix [4-9].
Motivated by several results obtained in random low rank approximation
theory, here we investigate the problem of image recovery from a small
number of random and noisy measurements, and we show that this problem is
equivalent to a randomized approximation of the null subspace of the
frequency response matrix.

\section{Frequency response matrix}

We consider a system consisting of an array of $N$ transceivers (i.e. each
antenna is an emitter and a receiver) located at $x_{n}\in R^{D}$ $%
(n=1,...,N)$, and a collection of $M$ distinct scatterers (targets) with
scattering coefficients $\rho _{m}$, located at $y_{m}\in R^{D}$ $(m=1,...,M)
$ (Fig. 1). Here, $D=1,2,3$ is the dimensionality of the space. Also, we
assume that the wave propagation is well approximated in the space-frequency
domain $(x,\omega )$ by the inhomogeneous Helmholtz equation [1-9]: 
\begin{equation}
\left[ \nabla ^{2}+k_{0}^{2}\eta ^{2}(x)\right] \psi (x,\omega )=-s(x,\omega
),
\end{equation}
where $\psi (x,\omega )$ is the wave amplitude produced by a localized
source $s(x,\omega )$, $k_{0}=2\pi \omega /c_{0}=2\pi /\lambda $ is the
wavenumber of the homogeneous background, with $\omega $ the frequency, $%
c_{0}$ the homogeneous background wave speed, and $\lambda $ the wavelength.
Here, $\eta (x)$ is the index of refraction: $\eta (x)=c_{0}/c(x)$, where $%
c(x)$ is the wave speed at location $x$. In the background we have $\eta
_{0}^{2}(x)=1$, while $\eta ^{2}(x)=1+\alpha (x)$, measures the change in
the wave speed at the scatterers location.

The fundamental solutions, or the Green functions, for this problem satisfy
the following equations: 
\begin{equation}
\left[ \nabla ^{2}+k_{0}^{2}\right] G_{0}(x,x^{\prime })=-\delta
(x-x^{\prime }),
\end{equation}
\begin{equation}
\left[ \nabla ^{2}+k_{0}^{2}\eta ^{2}(x)\right] G(x,x^{\prime })=-\delta
(x-x^{\prime }),
\end{equation}
for the homogeneous and inhomogeneous media, respectively. The fundamental
solution $G(x,x^{\prime })$ for the inhomogeneous medium can be written in
terms of that for the homogeneous one $G_{0}(x,x^{\prime })$ as: 
\begin{equation}
G(x,x^{\prime })=G_{0}(x,x^{\prime })+k_{0}^{2}\int \alpha
(z)G_{0}(x,z)G(z,x^{\prime })dz.
\end{equation}
This is an implicit integral equation for $G(x,x^{\prime })$. Since the
scatterers are assumed to be pointlike, the regions with $\alpha (z)\neq 0$
are assumed to be finite, and included in compact domains $\Omega _{m}$
centered at $y_{m}$, $m=1,...,M$, which are small compared to the wavelength 
$\lambda $. Therefore we can write: 
\begin{equation}
\alpha (z,\omega )=\sum_{m=1}^{M}\rho _{m}(\omega )\delta (z-y_{m}),
\end{equation}
and consequently we obtain: 
\begin{equation}
G(x,x^{\prime })\simeq G_{0}(x,x^{\prime })+\sum_{m=1}^{M}\rho _{m}(\omega
)G_{0}(x,y_{m})G(y_{m},x^{\prime }).
\end{equation}
If the scatterers are sufficiently far apart we can neglect the
multiple scattering among the scatterers $(G(y_{m},x^{\prime })\simeq
G_{0}(y_{m},x^{\prime }))$ and we obtain the Born approximation of the
solution [10]: 
\begin{equation}
G(x,x^{\prime })\simeq G_{0}(x,x^{\prime })+\sum_{m=1}^{M}\rho _{m}(\omega
)G_{0}(x,y_{m})G_{0}(y_{m},x^{\prime }).
\end{equation}
If $x$ corresponds to the receiver location $x_{i}$, and $x^{\prime }$
corresponds to the emitter location $x_{j}$, then we obtain: 
\begin{equation}
G(x_{i},x_{j})\simeq G_{0}(x_{i},x_{j})+H_{ij}(\omega ),
\end{equation}
where 
\begin{equation}
H_{ij}(\omega )=\sum_{m=1}^{M}G_{0}(x_{i},y_{m})\rho _{m}(\omega
)G_{0}(y_{m},x_{j}),\quad i,j=1,...,N,
\end{equation}
are the elements of the frequency response matrix $H(\omega )=[H_{ij}(\omega
)]$. The response matrix $H(\omega )$ is obviously a complex and symmetric $%
N\times N$ matrix, since the same Green function is used in both the
transmission and the reception paths.

\section{Computational time-reversal imaging}

An important step in CTRI is to determine the frequency response matrix $%
H(\omega )$. This can be done by performing a series of $N$ experiments, in
which a single element of the array is excited with a suitable signal $s(\omega )$
and we measure the frequency response between this element and all the other
elements of the array [1-9]. In general, given the Green function $G_{0}(x,x^{\prime })$, 
the general solution to the Helmholtz equation is the convolution: 
\begin{equation}
\psi (x,\omega )=(G_{0}*s)(x,\omega )=\int G_{0}(x,x^{\prime })s(x^{\prime
},\omega )dx^{\prime }.
\end{equation}
Thus, if the $j$ antenna emits a signal $s_{j}(\omega )$ then, using the
convolution theorem in the Fourier domain, the field produced at the
location $r$ is $G_{0}(r,x_{j})s_{j}(\omega )$. If this field is incident on
the $m$-th scatterer, it produces the scattered field $G_{0}(r,y_{m})\rho
_{m}(\omega )G_{0}(r,x_{j})s_{j}$. Thus, the total wave field, due to a
pulse emitted by a single element at $x_{j}$ and scattered by the $M$
targets can be expressed as: 
\begin{equation}
\psi (r,\omega )=\sum_{m=1}^{M}G_{0}(r,y_{m})\rho _{m}(\omega
)G_{0}(y_{m},x_{j})s_{j}(\omega ).
\end{equation}
If this field is measured at the $i$-th antenna we obtain: 
\begin{equation}
\psi (x_{i},\omega )=\sum_{m=1}^{M}G_{0}(x_{i},y_{m})\rho _{m}(\omega
)G_{0}(y_{m},x_{j})s_{j}=H_{ij}(\omega )s_{j}(\omega ).
\end{equation}
\qquad \qquad \qquad 

In CTRI one forms the symmetric self-adjoint matrix [1-9]: 
\begin{equation}
K(\omega )=H^{*}(\omega )H(\omega )=\overline{H}(\omega )H(\omega ),
\end{equation}
where the star denotes the adjoint and the bar denotes the complex conjugate
($H^{*}=\overline{H}$, since $H$ is symmetric). $\overline{H}$ is the
frequency-domain version of a time-reversed response matrix, thus $K(\omega )
$ corresponds to performing a scattering experiment, time-reversing the
received signals and using them as input for a second scattering experiment.
Therefore, time-reversal imaging relies on the assumption that the Green
function can be always calculated.

As long as the number of transceivers exceeds the number of scatterers, $%
M<N$, the matrix $K(\omega )$ is rank deficient and it has only $M$ non-zero
eigenvalues, with the corresponding eigenvectors $v_{m}(\omega )$, $%
m=1,...,M $. When the scatterers are well resolved, the eigenvectors can be
back-propagated as $g^{T}(r,\omega )v_{m}(\omega )$, and consequently the
radiated wavefields focus at target locations. Thus, each eigenvector can be
used to locate a single scatterer. Here, $g(r,\omega )$ is the Green
function vector, which expresses the response at each array element due to a
single pulse emitted from $r$: 
\begin{equation}
g(r,\omega )=\left[ 
\begin{array}{llll}
G_{0}(x_{1},r,\omega ) & G_{0}(x_{2},r,\omega ) & ... & G_{0}(x_{N},r,\omega
)
\end{array}
\right] ^{T}.
\end{equation}

The above result does not apply to the case of poorly-resolved targets. In
this case, the eigenvectors of $K(\omega )$ are linear combinations of the
target Green function vectors $g(y_{m},\omega )$. Thus, back-propagating one
of these eigenvectors generates a linear combination of wavefields, each
focused on a different target location. The subspace-based algorithms, based
on the multiple signal classification (MUSIC) method, can be used in this
more general situation [7-9]. The signal subspace method assumes that the
number $M$ of point targets in the medium is lower than the number of
transceivers $N$, and the general idea is to localize multiple sources by
exploiting the eigenstructure and the rank deficiency of the response matrix 
$H(\omega )$.

The SVD of the matrix $K(\omega )$ is given by: 
\begin{equation}
K(\omega )=\sum_{n=1}^{N}\lambda _{n}(\omega )u_{n}(\omega )v_{n}^{*}(\omega
),
\end{equation}
where $u_{n}(\omega )$ and $v_{n}(\omega )$ are the left and right singular
vectors. Since $H(\omega )$ is rank-deficient, all but the first $M$
singular values vanish: $\lambda _{1}(\omega )\geq ...\geq \lambda
_{M}(\omega )>0$, $\lambda _{j}(\omega )=0$, $j=M+1,...,N$. Therefore, the
first $M$ singular vectors span the essential signal subspace, while the
remaining $N-M$ columns span the null-subspace. The projection on the
null-subspace is given by: 
\begin{equation}
P_{null}(\omega )=\sum_{n=M+1}^{N}v_{n}(\omega )v_{n}^{*}(\omega
)=I-\sum_{n=1}^{M}v_{n}(\omega )v_{n}^{*}(\omega ),
\end{equation}
where $I$ is the identity matrix. It follows immediately that $%
P_{null}(\omega )K(\omega )=0$, and therefore $P_{null}(\omega )g(r,\omega
)=0$, for any $\omega $. Therefore, the target locations must correspond to
the peaks in the MUSIC pseudo-spectrum for any $\omega $: 
\begin{equation}
S_{MUSIC}(r,\omega )=\left\| P_{null}(\omega )g(r,\omega )\right\| ^{-2},
\end{equation}
where $g(r,\omega )$ is the free-space Green function vector. Thus, one can
form an image of the scatterers by plotting, at each point $r$, the quantity 
$S_{MUSIC}(r,\omega )$. The resulting plot will have large peaks at the
locations of the scatterers.

\section{Randomized null-subspace approximation}

Let us first present several results on random rank approximation. The SVD
of a $M\times N$ ($N\leq M$) matrix $A=[a_{ij}]$ can be written as [11]: 
\begin{equation}
A=\sum_{r=1}^{R}\lambda _{r}u_{r}v_{r}^{*},
\end{equation}
where $R\leq N$ is the rank of $A$, $u_{r}$ and $v_{r}$ are the left and
right singular vectors, and the singular values (in decreasing order) are: $%
\lambda _{1}\geq \lambda _{2}\geq ...\geq \lambda _{R}>0$. If $N>M$ one can
always compute the SVD of the transpose matrix and then swap the left and
right singular vectors in order to recover the SVD of the original matrix.
We also remind that the Frobenius and the spectral norms of $A$ are [11]: 
\begin{equation}
\left\| A\right\| _{F}=\sqrt{\sum_{m=1}^{M}\sum_{n=1}^{N}\left|
a_{mn}\right| ^{2}},\quad \left\| A\right\| _{2}=\lambda _{1}.
\end{equation}
If we define 
\begin{equation}
A_{K}=\sum_{k=1}^{K}\lambda _{k}u_{k}v_{k}^{*},
\end{equation}
for any $K\leq R$, then, by the Eckart-Young theorem, $A_{K}$ is the best
rank $K$ approximation to $A$ with respect to the spectral norm and the
Frobenius norm [11]. Thus, for any matrix $B$ of rank at most $K$, we have: 
\begin{equation}
\left\| A-A_{K}\right\| _{F}^{2}\leq \left\| A-B\right\| _{F}^{2},\quad
\left\| A-A_{K}\right\| _{2}^{2}\leq \left\| A-B\right\| _{2}^{2}.
\end{equation}
From basic linear algebra we have: 
\begin{equation}
A_{K}=A\left[\sum_{k=1}^{K}v_{k}v_{k}^{*}\right] .  
\end{equation}
Also, we say that a matrix $A$ has a good rank $K$ approximation if $A-A_{K}$
is small with respect to the spectral norm and the Frobenius norm.

Our problem is to substitute $A_{K}$ with some other rank $K$ matrix $D$ ,
which is much simpler than $A_{K}$, and does not require the full knowledge
of $A$. Therefore, the matrix $D$ must satisfy the general condition: 
\begin{equation}
\left\| A-D\right\| _{F}^{2}\leq \left\| A-A_{K}\right\| _{F}^{2}+\xi ,
\end{equation}
where $\xi $ represents a tolerable level of error for the given
application. Several important results have been recently obtained regarding
this problem.

It has been shown that one can compute a rank $K$ approximation of $A$ from
a randomly chosen submatrix of $A$ [12, 13]. For any $K\leq R$ and $%
0<\varepsilon ,\delta <1$ this method uses a matrix $D$, containing only a random
sample of $K$ rows of matrix $A$, so that: 
\begin{equation}
\left\| A-D\right\| _{F}^{2}\leq \left\| A-A_{K}\right\|
_{F}^{2}+\varepsilon \left\| A\right\| _{F}^{2},
\end{equation}
holds with probability of at least $1-\delta $. Recently, the above result
has been improved: 
\begin{equation}
\left\| A-D\right\| _{F}^{2}\leq (1+\varepsilon )\left\| A-A_{K}\right\|_{F}^{2} ,
\end{equation}
by taking into account that the additive error $\varepsilon \left\| A\right\| _{F}^{2}$ 
can be arbitrarily large compared to the true error $\left\| A-A_{K}\right\| _{F}^{2}$ [14]. 
These results show that the sparse matrix $D$ recovers almost as much from $A$ as
the best rank approximation matrix $A_{K}$.

In a different approach [15], it has been shown that one can substitute $%
A_{K}$ with a sparse matrix $D=[d_{ij}]$, where $d_{ij}=a_{ij}$ with
probability $p$, and $d_{ij}=0$ with probability $1-p$. This result asserts
that it is possible to find a good low rank approximation to $A$ even after
randomly omitting many of its entries. In particular, it has been shown that
the stronger the spectral features of $A$ the more of its entries we can
afford to omit.

Another observation is related to the noise effect on low rank
approximation. One can model this by adding to $A$ a matrix $F$ whose
entries are independent Gaussian random variables with mean $0$ and standard
deviation $\sigma $. As long as $\sigma $ is not too big, the optimal rank $K
$ approximation $(A+F)_{K}$ to $A+F$ will approximate $A$ nearly as well as $%
A_{K}$ [15]. This stability of low rank approximations with respect to
Gaussian noise is well-understood, and in fact low rank approximations are
frequently used with the explicit purpose of removing Gaussian noise.

The above results can be transferred to the CTRI problem by substituting the
matrix $A$ with the frequency response matrix $H(\omega )$, and considering
a sparse matrix $\widetilde{H}(\omega )$ satisfying the above conditions for
the matrix $D$. Also, from the CTRI considerations, we can form the matrix: 
\begin{equation}
\widetilde{K}(\omega )=\widetilde{H}^{*}(\omega )\widetilde{H}(\omega ),
\end{equation}
as a substitute for $K(\omega )$. Since $\widetilde{H}(\omega )$ is a rank $M
$ approximation of $H(\omega )$, the rank of $\widetilde{K}(\omega )$ will
also be $M$, and its SVD will be given by: 
\begin{equation}
\widetilde{K}(\omega )=\sum_{n=1}^{N}\mu _{n}(\omega )a_{n}(\omega
)b_{n}^{*}(\omega ),
\end{equation}
where $a_{n}(\omega )$ and $b_{n}(\omega )$ are the left and right singular
vectors, and all but the first $M$ singular values vanish: $\mu _{1}(\omega
)\geq ...\geq \mu _{M}(\omega )>0$, $\mu _{j}(\omega )=0$, $j=M+1,...,N$.
Now, since $\widetilde{H}(\omega )$ is a rank $M$ approximation of $H(\omega
)$, the last $N-M$ right singular vectors of $\widetilde{K}(\omega )$ will
approximate the null-subspace of $\widetilde{K}(\omega )$ (and implicitly of 
$\widetilde{H}(\omega )$). Thus, the approximate projection on the
null-subspace is 
\begin{equation}
\widetilde{P}_{null}(\omega )=\sum_{n=M+1}^{N}b_{n}(\omega )b_{n}^{*}(\omega
)=I-\sum_{n=1}^{M}b_{n}(\omega )b_{n}^{*}(\omega ),
\end{equation}
with the corresponding MUSIC pseudo-spectrum given by: 
\begin{equation}
\widetilde{S}_{MUSIC}(r,\omega )=\left\| \widetilde{P}_{null}(\omega
)g(r,\omega )\right\| ^{-2}.
\end{equation}

In order to illustrate and validate numerically the above results, we have
considered a two dimensional scenario, including $N=100$ transceivers,
separated by $d=\lambda /2$ and located at $x_{n}=\left[ 
\begin{array}{ll}
0 & n\lambda /2+a/2-N\lambda /4
\end{array}
\right] ^{T}$, where $a=100\lambda $ is the side of the imaging area. The
number of targets (with the scattering coefficients $\rho _{m}=1$) is set to 
$M=5$ and their position is randomly generated in the imaging area. The
computational image grid is also set to $L\times L=300\times 300$ pixels.
The two dimensional Green function is $G_{0}(x,x^{\prime })=\frac{i}{4}%
H_{0}^{(1)}(k_{0}|x-x^{\prime }|)$, where $H_{0}^{(1)}(.)$ is the zero order
Hankel function of the first kind. The noise level is characterized by the
signal to noise ratio (SNR). SNR compares the level of a desired signal to
the level of background noise. The higher the ratio, the less obtrusive the
background noise is. SNR measures the power ratio between a signal and the
background noise: 
\begin{equation}
SNR=P_{signal}/P_{noise}=(A_{signal}/A_{noise})^{2},
\end{equation}
where $P$ is average power and $A$ is root mean square (RMS) amplitude.

Let us first consider the case when the matrix $\widetilde{H}(\omega )$ is
obtained by randomly selecting $M\leq J\leq N$ rows of the matrix $H(\omega )
$. In Figure 2 we give the results obtained for the extreme case of $J=M$
and for different levels of noise, $SNR=\infty ,80,40,20,10,5,2,1$. One can
see that even for this extreme case the results are actually pretty good. By
increasing $J$ the quality of the image improves even at high levels of
noise, as shown on Figure 3, where $J=1M,2M,3M,4M$ and the noise is fixed at 
$SNR=2$ on the first line of images, and respectively $SNR=1$ on the second
line of images. If $J<M$ the algorithm doesn't work.

In Figure 4 we give the results obtained when the elements of the matrix $%
\widetilde{H}(\omega )$ are selected randomly as: $\widetilde{H}_{ij}(\omega
)=H_{ij}(\omega )$ with probability $p$, and $\widetilde{H}_{ij}(\omega )=0$
with probability $1-p$ (we also conserved the symmetry 
$\widetilde{H}_{ij}(\omega )=\widetilde{H}_{ji}(\omega )$ during the random
selection). The figure 'matrix' is organized on lines and columns. The
lines correspond to the probability $p=1,0.5,0.25,0.1$, and the columns
correspond to the noise level $SNR=\infty ,10,5,2$.

\section{Conclusion}

We have shown that the problem of image recovery from a small number of
random and noisy measurements is equivalent to a randomized approximation of
the null subspace of the frequency response matrix. The obtained results
show that one can recover the sparse time-reversal image from fewer (random)
measurements than conventional methods use. From the analytical results and
the numerical experiments we conclude that the minimum number of
measurements is $MN\ll N^{2}$, where $M$ is the rank of the full matrix $%
H(\omega )$.

\pagebreak

\clearpage
\begin{figure}
\centering
\includegraphics[width=7cm]{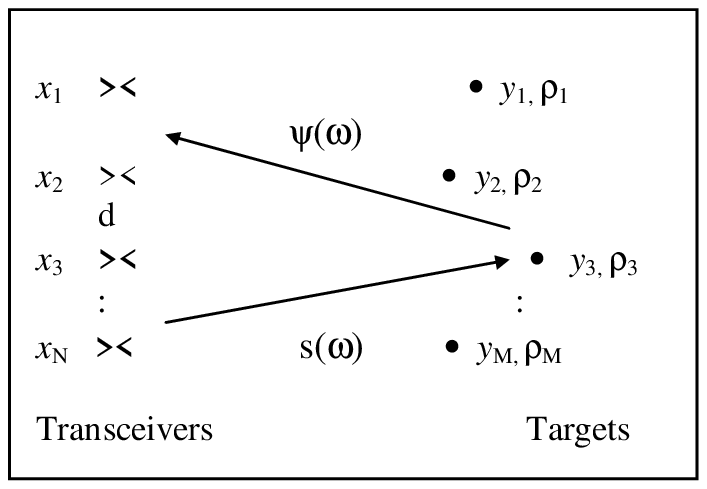}
\caption{\label{Fig1} Geometry of a time-reversal imaging experiment, containing $N$
transceivers and $M$ scattering targets.}
\end{figure}

\clearpage
\begin{figure}
\centering
\includegraphics[width=12cm]{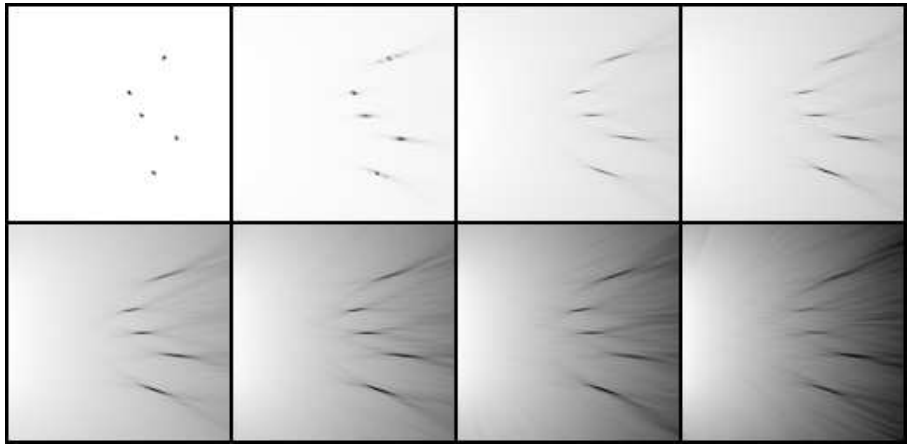}
\caption{\label{Fig2} Numerical results for $\widetilde{H}(\omega )$ obtained by randomly
selecting $J=M$ rows of the matrix $H(\omega )$, for different levels of
noise: $SNR=\infty ,80,40,20,10,5,2,1$ (from top left corner to bottom
right corner).}
\end{figure}

\clearpage
\begin{figure}
\centering
\includegraphics[width=12cm]{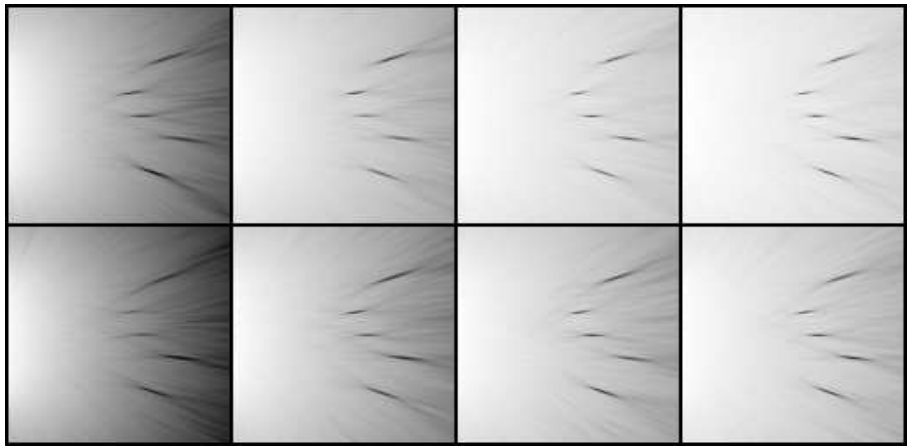}
\caption{\label{Fig3} Numerical results for $\widetilde{H}(\omega )$ obtained by randomly
selecting $J=M,2M,3M,4M$ rows of the matrix $H(\omega )$: first line $SNR=2$; second line $SNR=1$.}
\end{figure}

\clearpage
\begin{figure}
\centering
\includegraphics[width=12cm]{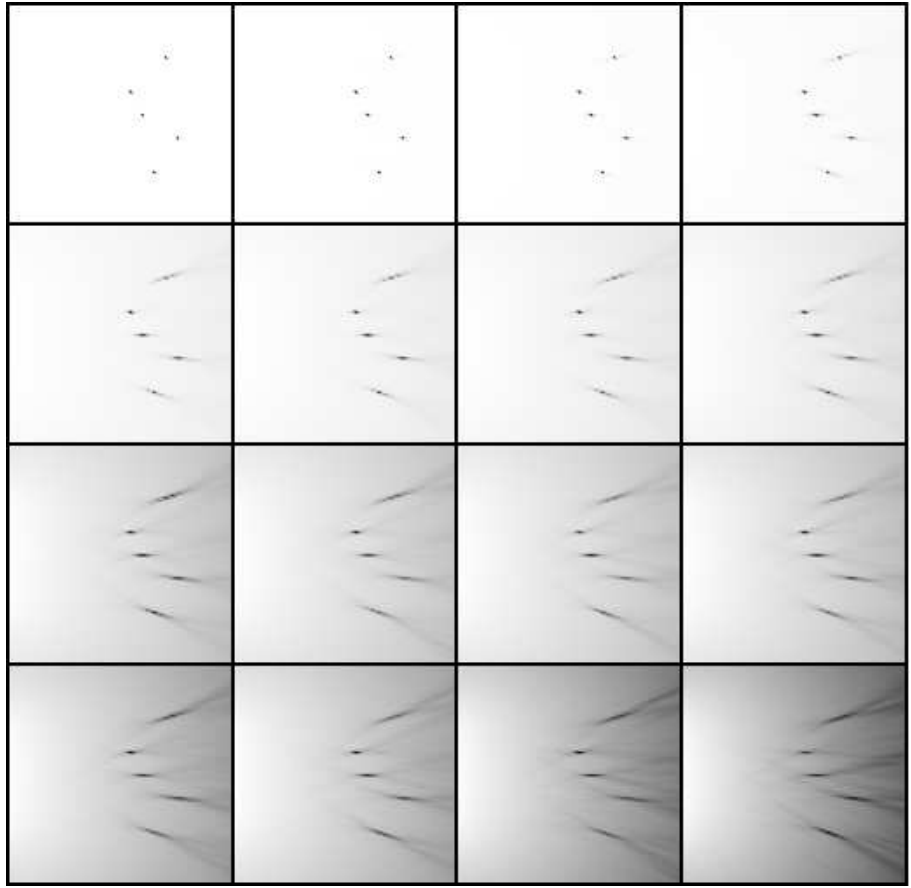}
\caption{\label{Fig4} Numerical results obtained when the elements of the matrix $%
\widetilde{H}(\omega )$ are selected randomly as: $\widetilde{H}_{ij}(\omega
)=H_{ij}(\omega )$ with probability $p$, and $\widetilde{H}_{ij}(\omega )=0$
with probability $1-p$. The lines correspond to the probability $%
p=1,0.5,0.25,0.1$, and the columns correspond to the noise level $SNR=\infty
,10,5,2$.}
\end{figure}

\end{document}